\documentclass[%
reprint,
 amsmath,amssymb,
 aps,
pra,
]{revtex4-2}

\usepackage{graphicx}
\usepackage{dcolumn}
\usepackage{bm}
\usepackage{hyperref}
\usepackage[titletoc,title]{appendix}

\begin{document}

\preprint{APS/123-QED}

\title{Angular dependency of spatial frequency modulation in diffusion media}

\author{Yun Chen}
\author{Chengyuan Wang}
\email{wcy199202@gmail.com}
\author{Zibin Jiang}
\author{Wei Zhang}
\author{Zehao Shen}
\author{Hong Gao}
\author{Fuli Li}
\address{Ministry of Education Key Laboratory for Nonequilibrium Synthesis and Modulation of Condensed Matter, Shaanxi Province Key Laboratory of Quantum Information and Quantum Optoelectronic Devices, School of Physics, Xi'an Jiaotong University, Xi'an 710049, China}
%

\date{\today}

\begin{abstract}
An optical field will undergo coherent diffusion when it is mapped into thermal-motioned atoms, e.g., in a slow or storage light process. As was demonstrated before, such diffusion effect is equivalent to a spatial low-pass filter attenuating the high spatial frequency (SF) components of the optical field. Here, employing electromagnetically induced transparency (EIT) based light storage in hot atomic vapor, we demonstrate that the angular deviation between the control and probe beams could be utilized as a degree of freedom to modulate the SF of the probe beam. The principle is to change the diffusion-induced low-pass filter into a band-pass filter, whose SF response can be tuned by varying the direction and magnitude of the angular deviation. Transverse multimode light fields, such as optical images and Laguerre-Gaussian modes are utilized to study such SF modulation. Our findings could be broadly applied to the fields of quantum information processing, all-optical image manipulation and imaging through diffusive media.

\end{abstract}

\maketitle


\section{introduction}
Coherent manipulation of transverse multimode light fields, such as optical modes with representative transverse structures \cite{doi:10.1116/5.0016007,ding2013single,de2014off,Yu:21} or arbitrary optical images \cite{PhysRevA.87.053830,PhysRevLett.100.223601,PhysRevLett.100.123903,doi:10.1063/5.0053849}, via light-atom interaction is essential for the development of quantum optics and quantum information. External light or magnetic fields can modulate the optical response of the atoms, then change the absorption and dispersion of the optical field propagating through such media. For instance, under the electromagnetically induced transparency (EIT) condition, the susceptibility of the atomic media could be modulated through a control field, rendering the variation of the absorption \cite{PhysRevLett.114.123603,lukin2001controlling}, diffraction \cite{PhysRevLett.102.043601,firstenberg2009elimination}, deflection \cite{PhysRevLett.103.033003,PhysRevA.81.063824}, or group velocity \cite{PhysRevLett.84.5094,hau1999light} of a probe field. Thermal alkali atomic ensemble, possessing the advantages of low cost and large coupling strength when interacting with light, has always been a preferable media to conduct relevant researches \cite{Wang:18,Yang:19,PhysRevLett.98.203601}. 

A distinctive characteristic of hot atoms is the continuous thermal motion, which would change the transverse structure or propagation dynamic of a light field when it is slowed or stored \cite{PhysRevA.77.043830}. For instance, Firstenberg \emph{et al.} demonstrated that a light field slowed in hot atoms would be subject to a new form of diffraction, referred to as motional-induced diffraction (MID) \cite{RevModPhys.85.941}. By manipulating the EIT linear susceptibility in momentum space, MID could eliminate the paraxial diffraction of optical images throughout their propagation \cite{firstenberg2009elimination}. Another prominent motional-induced effect is coherent diffusion. In the general light storage experiments associated with collinear probe-control configuration, coherent diffusion operates as a low-pass filter in the spatial frequency (SF) space ($\textbf{q}$-space) and wipes out the high SF components of the stored probe field \cite{Chriki:19,Smartsev:20,wang21}. For a scalar field or an optical image, the details of its spatial structure depend on the high SF components, and the loss of which would result in a blurry image. This is a defect that degrades the image's quality when it is manipulated with the slow and storage light techniques in the hot vapor cell. To date, the conventional approaches to solve these problems are mostly based on modulating the wavefront of the optical image \cite{PhysRevLett.100.223601,doi:10.1063/5.0053849,PhysRevLett.100.123903,PhysRevA.86.013844}. For instance, loading an appropriate phase to the image according to the optical phase-shift lithography principle \cite{PhysRevLett.100.223601,doi:10.1063/5.0053849}, storing the Fourier transform of the image \cite{PhysRevLett.100.123903}, or encoding images into thermal light and adopting the correlation imaging technique \cite{PhysRevA.86.013844} could all alleviate the influence of atomic diffusion on the quality of retrieved images. Nevertheless, these methods need additional optical devices such as a spatial light modulator (SLM) for wavefront modulation, which limits the flexibility of their practical applications. 

In this paper, we demonstrate that in a light storage experiment where the control and probe beams propagate non-collinearly, the angular deviation between the two beams could be utilized as a degree of freedom (DOF) to modulate the transverse structure of multimode light fields. The principle is to change the diffusion-induced low-pass filter into a band-pass filter, whose SF response can be tuned by varying the direction and magnitude of the angular deviation, as predicted by our theoretical model. We first investigate the modulation effect of the band-pass filter on several spatially periodic patterns, including the Laguerre-Gaussian (LG) modes. Then we demonstrate that for a general optical image whose SF mainly distributes along a certain direction, one can improve its storage performance by designing an appropriate band-pass filter. The method used here has great superiority and feasibility over the conventional approaches as no additional devices are required for wavefront modulation of the probe beam. The above demonstrations will be useful for multimode quantum memory, all-optical structured field manipulation, imaging through diffusive media, etc.

\section{Theoretical  model}
\begin{figure}[tp]
\centering\includegraphics[width=\linewidth]{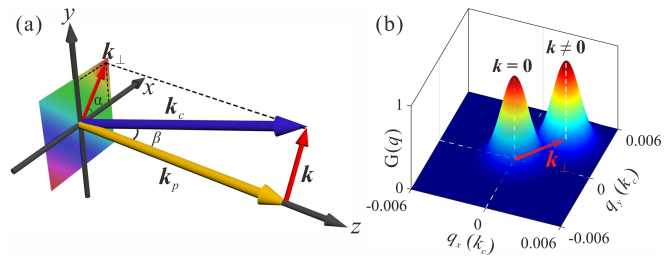}
\caption{(a) Geometry of the wavevectors of control and probe beams crossing the vapor cell with the center at $z=0$. (b) Spatial frequency response of collinear geometry (${\textbf k} = 0$) and noncollinear geometry (${\textbf k} \neq 0$). Here, the diffusion coefficient and storage time are $D$ = 25 cm/s$^2$ and $t$ = 3 $\mu$s, respectively.}\label{fig:1}
\end{figure}

To study the angular dependency of SF modulation, we consider a typical storage process in thermal atomic vapor with the control and probe fields constituting a $\Lambda$-type EIT structure. The probe has a wavevector $\textbf{{k}}_p$ ($|\textbf{k}_p|$ =2$\pi/ \lambda_{p}$ with $\lambda_{p}$ being the wavelength of the probe), and is encoded with transverse images. By turning off the control beam with wavevector $\textbf{{k}}_c$ ($|$$\textbf{k}_c$$|$ = 2$\pi/ \lambda_{c}$ with $\lambda_{c}$ the wavelength of the control), both the spatial structure of the probe beam $\psi(\textbf{r}, 0)$ and two beams' wavevector difference $\textbf{k}=\textbf{k}_{c}-\textbf{k}_{p}$ are mapped onto the atomic ground state coherence $\rho_{12}({\textbf r}, 0)$ \cite{liu2001observation,PhysRevA.93.063819}. For simplicity, we only consider diffusion in the transverse plane ($x-y$ plane). The dynamics of $\rho_{12}({\textbf r}, 0)$ during storage duration $t$ are governed by \cite{RevModPhys.85.941}
\begin{equation}
\partial_{t}\rho_{12}({\textbf r}, t)-D(\nabla_{\perp}-i \textbf{k}_{\perp})^2 \rho_{12}({\textbf r}, t)=0,
\label{Eq. 1}
\end{equation}
where $D$ is the spatial diffusion coefficient proportional to $T^{3/2}/P$ \cite{PhysRevA.79.042905} with $T$ the temperature of the atomic vapor and $P$ the pressure of the buffer gas, $\textbf{k}_{\perp}$ is the projection of $\textbf{k}$ in $x-y$ plane (see Fig. \ref{fig:1}(a)) and its components in the $x$ axis and the $y$ axis are labeled $\textbf{k}_{x}$ and $\textbf{k}_{y}$, respectively. 

Since the retrieved probe field $\psi(\textbf{r}, t)\propto \rho_{12}({\textbf r}, t)$, Eq. (\ref{Eq. 1}) can be rewritten in the form of  $\psi(\textbf{r}, t)$ and further transformed into a representation in Fourier space
\begin{equation}
\partial_{t}\tilde{\psi}({\textbf q}, t)+D(\textbf{q}-\textbf{k}_{\perp})^2 \tilde{\psi}({\textbf q}, t)=0,
\label{Eq. 2}
\end{equation}
where $\tilde{\psi}({\textbf q}, t)$ is the Fourier transform (FT) of $\psi(\textbf{r}, t)$ in $\textbf{r}$. 

The solution of Eq. (\ref{Eq. 2}) is
\begin{equation}
\tilde{\psi}({\textbf q}, t)=\tilde{\psi}({\textbf q}, 0)e^{-Dt(\textbf{q}-\textbf{k}_{\perp})^{2}},
\label{Eq. 3}
\end{equation}
where $\tilde{\psi}({\textbf q}, 0)$ is the FT of $\psi(\textbf{r}, 0)$ in $\textbf{r}$ space.

We observe that, when $\textbf{k}_{\perp}=0$, Eq. (\ref{Eq. 3}) can be simplified into $\tilde{\psi}({\textbf q}, t)=\tilde{\psi}({\textbf q}, 0)e^{-Dt\textbf{q}^{2}}$, indicating that coherent diffusion operates as a Gaussian low-pass filter, whose SF response can be expressed as $G(\textbf{q})=e^{-Dt\textbf{q}^{2}}$. For an optical image that passes such a filter, its high SF components will suffer orientation-independent attenuation, resulting a loss of overall detail in the image. However, when $\textbf{k}_{\perp}\neq0$, the center of the low-pass filter corresponding to $\textbf{k}_{\perp} = 0$ will move from $\textbf{q}=0$ to $\textbf{q}=\textbf{k}_{\perp}$, and the altered SF response $G(\textbf{q})=e^{-Dt(\textbf{q}-\textbf{k}_{\perp})^{2}}$ has the largest transmissivity at $\textbf{q}=\textbf{k}_{\perp}$, as shown in Fig. \ref{fig:1}(b). As a result, the low-pass filter is converted into a Gaussian band-pass filter with the center SF located at $\textbf{k}_{\perp}$ and the bandwidth proportional to $\frac{1}{Dt}$. In this situation, the influence of diffusion on the optical image will become ``directional'' and controllable, since the high SF components located around $\textbf{k}_{\perp}$ are able to be maintained and meanwhile $\textbf{k}_{\perp}$ can be flexibly customized by regulating the incident angle of the control beam.

\section{Experiment results and discussions}
To investigate the influence of $\textbf{k}_{\perp}$ on the evolution of the beam's profile, we perform EIT-based light storage experiments in a thermal $^{87}$Rb atomic vapor cell. The cell contains 8 Torr neon buffer gas and is heated to 60$^{\circ}$C, rendering a diffusion coefficient of $D \approx 25$ cm$^{2}$/s \cite{Chen:21}. The adopted experimental setup (see Appendix A for details) is similar to that in our previous work \cite{doi:10.1063/5.0053849}. Here the wavevector of the control beam is modulated while the probe's wavevector remains unchanged and they intersect at the center of the atomic cell. The geometry of their wavevectors crossing the vapor cell is illustrated in Fig. \ref{fig:1}(a). The angle between $\textbf{k}_{c}$ and $\textbf{k}_{p}$ is marked as $\beta$, and the angle between $\textbf{k}_{\perp}$ and $x$ axis is denoted as $\alpha$. Since $\beta$ is small (a few milliradians), and k$_{c}$ and k$_{p}$ are $10^{4}$ to $10^{5}$ times larger than k$_{\perp}$ in practical experiment, we use the approximation of $\textbf{k}_{\perp}\approx\textbf{k}\approx \beta \textbf{k}_{c}$ for theoretical predictions.

\subsection{Storage of multimode transverse images: exploring angular dependency of the spatial frequency modulation}
\begin{figure}[tp]
\centering\includegraphics[width=8cm]{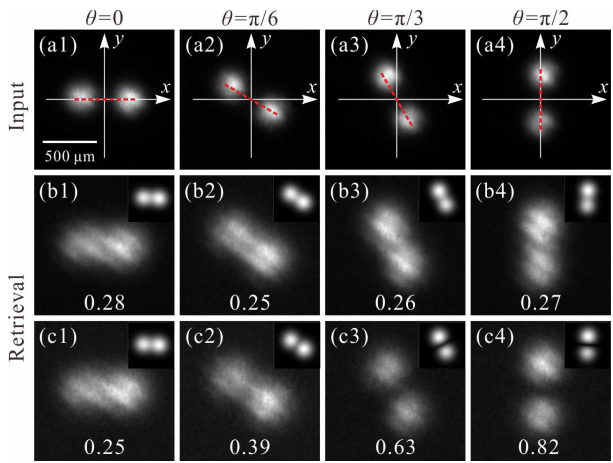}
\caption{Storage of double-petal images. (a1)-(a4) Input images with different rotation angles $\theta$ indicated by red dotted lines. (b1)-(b4) and (c1)-(c4) Retrieved images stored for 3 $\mu$s under collinear [(b1)-(b4)] and noncollinear [(c1)-(c4)] configurations. For the noncollinear case, the control and probe are separated by an angle of $\beta$ = 1.48 mrad along $y$ axis ($\alpha$ = $\pi/2$). The white numbers indicate the visibilities of the images' center. The illustrations in the upper right corners are theoretical simulations.}\label{fig:2}
\end{figure}

We begin by storing four double-petal images with the storage time fixed at 3 $\mu$s, and investigate the influence of the direction and magnitude of the angular deviation on the stored images. The petals (each has a waist radius of 100 $\mu$m) are separated by 475 $\mu$m and rotated clockwise about the origin of the Cartesian coordinates by angles of $\theta$ = 0, $\pi$/6, $\pi$/3, and $\pi$/2, as shown in Fig. \ref{fig:2}(a1)-(a4). First, the control and probe beams are set to collinear propagation (namely, $\textbf{k}_{\perp}=0$) and the retrieved images are shown in Fig. \ref{fig:2}(b1)-(b4). It is quite clear that in this case coherent diffusion exhibits isotropic features, which embodies in the fusion of the two petals irrespective of the rotation angle $\theta$. Next, the two beams are separated by an angle of $\beta$ = 1.48 mrad along $y$ axis ($\alpha$ = $\pi/2$), and the corresponding storage results are shown in Fig. \ref{fig:2}(c1)-(c4). It can be seen that as $\theta$ rotates more toward $\pi/2$, the double-petals are less affected by coherent diffusion and the crosstalk between the petals are getting less prominent. In other words, in this case coherent diffusion exhibits anisotropic and $\textbf{k}_{\perp}$ dependence. The corresponding theoretical results, obtained by taking the inverse FT of Eq. (3), are also presented in the insets of Fig. 2 and found to agree well with the experimental results.

To quantify the influence of diffusion on the retrieved patterns, we extract the center points of the images and calculate their visibilities (shown in the white numbers in Fig. \ref{fig:2}) with the background noise subtracted. Specifically, the visibility $V$ is calculated by $V=(I_{max}-I_c)/(I_{max}+I_c)$, where $I_c$ and $I_{max}$ are the intensity of the image center and the maximum intensity of the double-petal, respectively. The subtraction of the background noise is to reduce the influence of some other factors on the calculated visibility, such as the lower storage efficiency caused by the increased angle deviation. Evidently, when $\textbf{k}_{\perp}$ = 0, the visibility basically remains unchanged at a low value (about 0.26) as $\theta$ increases from 0 to $\pi/2$. In contrast, when $\textbf{k}_{\perp} \neq 0$, the visibility increases with the increase of $\theta$ and reaches a maximum (0.82) at $\theta = \pi/2$. The visibility of Fig. \ref{fig:2}(c4) improves by 0.55 compared with Fig. \ref{fig:2}(b4).

The above phenomenon can be interpreted in terms of the $\textbf{k}_{\perp}$-dependent spatial frequency modulation. It should be illustrated that one double-petal image's SF mainly distributes along the centerline of the two petals, as indicated by the red-dotted lines in Fig. \ref{fig:2}(a1)-(a4). Now let us discuss the case of $\textbf{k}_{\perp} = 0$ first. As we demonstrated earlier, in the absence of $\textbf{k}_{\perp}$, coherent diffusion acts as a Gaussian low-pass filter and attenuates the high SF components of the probe. Thus irrespective of $\theta$, the high SF components of the double-petal image will be filtered, which results in the blurring of the images, as seen in Fig. \ref{fig:2}(b1)-(b4). Next we analyze the case of $\textbf{k}_{\perp} \neq 0$. With the presence of $\textbf{k}_{\perp}$, the center of the low-pass filter's SF response will shift by $\textbf{k}_{\perp}$. When $\theta=\pi/2$, the SF spectrum of the double-petal image mainly distributes along $y$-axis, which is consistent with the orientation of $\textbf{k}_{\perp}$ (=$\textbf{k}_{y}$), thus facilitating the passage of the high SF components and ensuring a crisp outline of the image. As shown in Fig. \ref{fig:2}(c1)-(c3), the closer $\theta$ gets to $\pi/2$, the clearer of the image (higher visibility).

\begin{figure}[bp]
\centering\includegraphics[width=8cm]{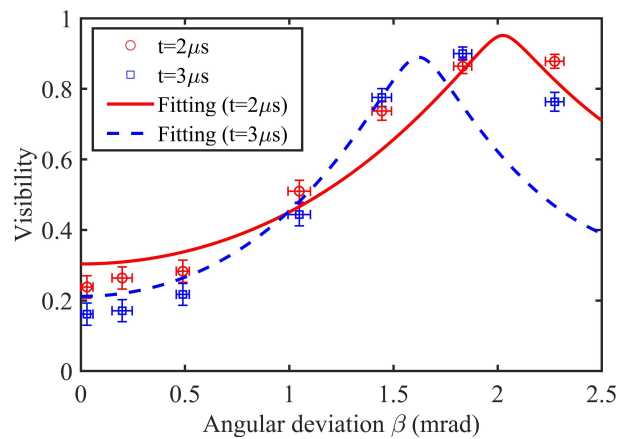}
\caption{The visibility of the double-petal image against the angle deviation $\beta$ between the control and probe beams. Here, $\theta$ is fixed at 0 [Fig. \ref{fig:2}(a1)] and $\textbf{k}_{\perp}$ is rotated along $x$-axis (i.e., $\alpha=0$). The solid and dashed curves are theoretical predictions. The experimental errors are due to the accuracy of determining the angular deviation and the detection noise.}\label{fig:3}
\end{figure}

The above is a qualitative explanation of this phenomenon, which can also be comprehended quantitatively from the following point of view. At time $t=0$, the probe photons are absorbed by the atoms and stored as spin waves (SWs). The wavevector mismatch ($\textbf{k}$) between the probe and the control beams is also imprinted onto the spin waves. At time $t=\tau$, the $i$-th atom with velocity $\textbf{v}^{i}$ will diffuse to a new place and the its SW will accumulate a $\textbf{k}$-related coefficient $e^{i\textbf{k}\cdot\textbf{v}^{j}\tau}$. At the same time, the $j$-th atom with velocity $\textbf{v}^{j}$ could diffuse to the same place and the corresponding coefficient of its SW is $e^{i\textbf{k}\cdot\textbf{v}^{j}\tau}$. If the phase difference between the two SWs satisfies 
\begin{equation}
\textbf{k}\cdot(\textbf{v}^{i}-\textbf{v}^{j})\tau =\pi,
\label{Eq. 4}
\end{equation}
then when the two SWs are transferred back to the probe photons, they will interfere destructively. Here for simplicity we also only consider diffusion in the $x-y$ plane, thus Eq. (\ref{Eq. 4}) is simplified to 
\begin{equation}
\textbf{k}_{\perp}\cdot(\textbf{v}_{\perp}^{i}-\textbf{v}_{\perp}^{j})\tau =\pi,
\label{Eq. 5}
\end{equation}

We can infer from Eq. (\ref{Eq. 5}) that, when $\textbf{k}_{\perp}=0$, $\textbf{k}_{\perp}\cdot(\textbf{v}_{\perp}^{i}-\textbf{v}_{\perp}^{j})\tau \equiv 0$, therefore no destructive interference occurs irrespective of $\textbf{v}_{\perp}$, as demonstrated in Fig. \ref{fig:2}(b1)-(b4). In Fig. \ref{fig:2}(c1), although $\textbf{k}_{\perp}=\textbf{k}_{y}\neq 0$, diffusion along $x$-axis ($\textbf{v}_{\perp}=\textbf{v}_{x}$) will lead to $\textbf{k}_{y} \perp (\textbf{v}_{x}^{i}-\textbf{v}_{x}^{j})$, thus $\textbf{k}_{\perp}\cdot(\textbf{v}_{\perp}^{i}-\textbf{v}_{\perp}^{j})\tau =0$ and also no destructive interference occurs. 
 Except for the above two situations, for a given $\tau$, one can find the optimal $\textbf{k}_{\perp}$ to approach the destructive interference condition and increase the visibility of the retrieved image to the greatest extent, as in the case of Fig. \ref{fig:2}(c4). 

Equation (\ref{Eq. 5}) also implies that the visibility of the double-petal image and the angular deviation $\beta$ are not always positively correlated. To quantify this, we fix $\theta$ at 0 (see Fig. \ref{fig:2}(a1)) and orientate $\textbf{k}_{\perp}$ along $x$-axis (i.e., $\alpha=0$), and then measure the retrieved image's visibility against $\beta$ with $t = 2$ $ \mu$s and  $t = 3$ $ \mu$s. Both the experimental results (red circles and blue squares) and the theoretical simulations (solid and dashed lines) are depicted in Fig. \ref{fig:3}. As can be seen that the image's visibility is improved with the increase of $\beta$ and reaches the maximum, and then decreases due to the imperfect destructive interference induced by the overlarge angular deviation. In addition, the optimal $\beta$ for $t = 2$ $\mu$s is right moved compared with $t = 3$ $\mu$s. This is reasonable since for a smaller $t$, $\textbf{k}_{\perp}$ should be larger to satisfy Eq. (\ref{Eq. 5}).

 The principle above is similar to Ref. \cite{PhysRevLett.100.223601,doi:10.1063/5.0053849}, in which the authors increase the stored image's visibility by imposing $\pi$-phase shift between neighboring features of the image.  However, rotating the wavevector offers greater convenience since there is no need to elaborately design and load the phase pattern to the image \cite{PhysRevLett.100.223601,doi:10.1063/5.0053849}. It is also worth noting that Eq. (\ref{Eq. 1}) is derived by using the Boltzmann relaxation method [20], which already takes into account the thermal equilibrium Boltzmann distribution in velocity space. This shows that our scheme is also applicable in atomic vapor with dominant Doppler effect. Moreover, the essence of our proposal is that the wavevector difference written into the spin wave will lead to the phase difference of atoms at different positions, resulting in the interference effect, and then changing the spatial structure of the stored image. The Doppler effect mainly changes the atomic decoherence process, which may reduce the storage performance, but does not determine the occurrence of the interference.
\begin{figure}[tp]
\centering\includegraphics[width=8cm]{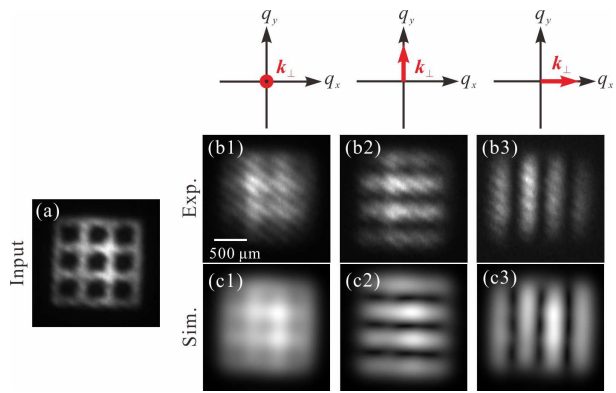}
\caption{Storage of grid patterns. (a) Input image. (b1)-(b3) Retrieved images stored for 3 $\mu$s under three cases of customized $\textbf{k}_{\perp}$ depicted the top sketches. Specific parameters used from left to right: $\alpha=0$ $\&$ $\beta \approx 0$ mrad, $\alpha=\pi/2$ $\&$ $\beta \approx 1.76$ mrad, $\alpha=0$ $\&$ $\beta \approx 1.76$ mrad. (c1)-(c3) Theoretical simulations for (b1)-(b3).}\label{fig:4}
\end{figure}

In addition, a more representative grid pattern, which has spatial frequency components along both $x$-axis and $y$-axis, is used to study the $\textbf{k}_{\perp}$ modulating effect. The input image is shown in Fig. \ref{fig:4}(a) and the retrieved images after 3 $\mu$s storage time are shown in Fig. \ref{fig:4}(b1)-(b3), while Fig. \ref{fig:4}(c1)-(c3) are the corresponding simulations. In Fig. \ref{fig:4}(b1) \& (c1),  $\textbf{k}_{\perp}=0$, hence atomic diffusion is isotropic and the grid merges together and becomes unrecognizable. In Fig. \ref{fig:4}(b2) \& (c2) and 4(b3) \& (c3), $\textbf{k}_{\perp}$ is along $y$-axis and $x$-axis ($\textbf{k}_{y}$ and $\textbf{k}_{x}$, both with $\beta \approx 1.76$ mrad). In these cases, the retrieved images are turned into a horizontal four-line ``$\rotatebox{90}{$||||$}$'' and a vertical four-line ``$||||$''. To explain this phenomenon more clearly, we can treat the grid as the superposition of ``$\rotatebox{90}{$||||$}$'' and ``$||||$''. Since the spatial frequency of ``$\rotatebox{90}{$||||$}$'' (``$||||$'') mainly distributes along $y$-axis ($x$-axis), $\textbf{k}_{y}$ ($\textbf{k}_{x}$) preserves the high-frequency components of ``$\rotatebox{90}{$||||$}$'' (``$||||$'') while isolates most of the frequency components of ``$||||$'' (``$\rotatebox{90}{$||||$}$''). Therefore, only ``$\rotatebox{90}{$||||$}$'' (``$||||$'') retains after storage with the presence of $\textbf{k}_{y}$ ($\textbf{k}_{x}$). The simulations based on Eq. (\ref{Eq. 3}) agree well with the experimental results, and the detailed atomic filtering process is given in Appendix B.

\subsection{Storage of Laguerre-Gaussian modes in the presence of wavevector misalignment}
\begin{figure}[tp]
\centering\includegraphics[width=8cm]{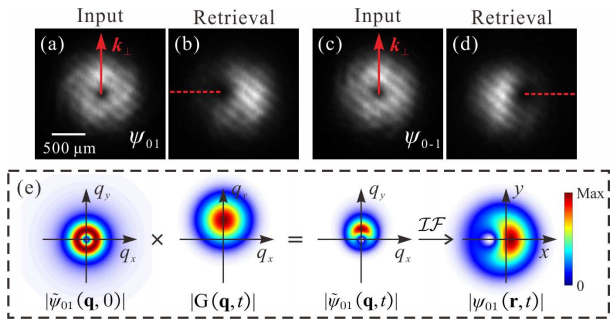}
\caption{Storage of the LG modes ($\Psi_{01}$ and $\Psi_{0-1}$) under the noncollinear configuration. (a)-(d) Input LG modes and the corresponding retrieved patterns with a storage time of $t=3$ $\mu$s. The red lines with arrows show the direction of the wavevector difference $\textbf{k}_{\perp}$, and the red dashed lines indicate the locations of the defects. (e) Theoretical diagram illustrating the cause of the defect.}\label{fig:5}
\end{figure}

In particular, we further investigate the effect of wavevector misalignment on the storage of Laguerre-Gaussian (LG) modes, which are widely used in quantum information processing \cite{yang2018multiplexed,PhysRevLett.98.203601,nicolas2014quantum}. Different from the above beams that are transverse images carrying no phase information, LG modes are known to carry vortex phases or orbital angular momentums (OAMs). Generally, an LG mode is characterized by two parameters, the radial index $p$ and the azimuthal index $l$. Here, we simply denote its amplitude field $\psi_{pl}(\textbf{r})$ as 
\begin{equation}
\psi_{pl}(\textbf{r}) = \psi_{pl}(r) {\rm{exp}}(il\phi),
\label{Eq. 6}
\end{equation}
where $ \psi_{pl}(r)$ is the Laguerre-Gaussian function and $\phi$ is the azimuth angle in the plane-polar coordinate. We employ two typical LG modes, $\psi_{01}$ and $\psi_{0-1}$, and store them for 3 $\mu$s with $\alpha = \pi/2$ and $\beta = 1.47$ mrad. The experimental results are shown in Fig. \ref{fig:5}(a)-(d). An interesting phenomenon is observed that the retrieved pattern has a defect, and the orientation of the defect is always perpendicular to $\textbf{k}_{\perp}$ and related to the sign of the topological charge.

This phenomenon can be analyzed in terms of the Fourier transform of the LG mode. To this end, by applying the 2D Fourier operator ($\mathcal{F}$) to $\psi_{pl}(\textbf{r})$, the representation of the LG mode in the Fourier space can be derived \cite{Yu_1998}
\begin{equation}
\widetilde{\psi}_{pl}(\textbf{q}) = \mathcal{F}[\psi_{pl}(\textbf{r})] = {\rm{exp}}(i p\pi)\widetilde{\psi}_{pl}(q){\rm{exp}}[il(\varphi+\pi/2)],
\label{Eq. 7}
\end{equation}
where $\varphi$ is the azimuth angle in the Fourier plane-polar coordinate. Regardless of $p$ and the beam size of the light field, it is easy to find from Eq. (\ref{Eq. 7}) that the Fourier-transformed LG mode maintains its original amplitude distribution and OAM charge except that its overall spiral phase rotates by $\pi/2$ and the rotation direction depends on the sign of $l$. Taking $\psi_{01}$ as an example, when $\textbf{k}_{\perp}$ is along the positive $y$-axis, the center of the filter's SF response will shift to the positive $y$-axis, resulting in a defect of $\widetilde{\psi}_{01}(\textbf{q},t)$ in the negative $y$-axis. Hence, after inverse Fourier transformed into real space, the defect rotates clockwise by $\pi/2$, as illustrated in Fig. \ref{fig:5}(e). These results are similar to that of Ref. \cite{Srinivas:18}, in which the authors observed that when a truncated vortex beam propagates to infinity, its truncated direction also rotates by $\pi/2$.

Our findings here provide a new interpretation mechanism for some imperfect experimental results, which were ignored in previous works \cite{Chriki:19,Smartsev:20}. In some scenarios where misalignment between the control and the probe is anticipated, such as spatially filtering the control noise from the weak probe beam when storing single-photon level quantum states \cite{wang2020efficient} in hot atomic vapor \cite{PhysRevLett.98.203601}, such diffusion influence should be taken into consideration.

\begin{figure}[tp]
\centering\includegraphics[width=8cm]{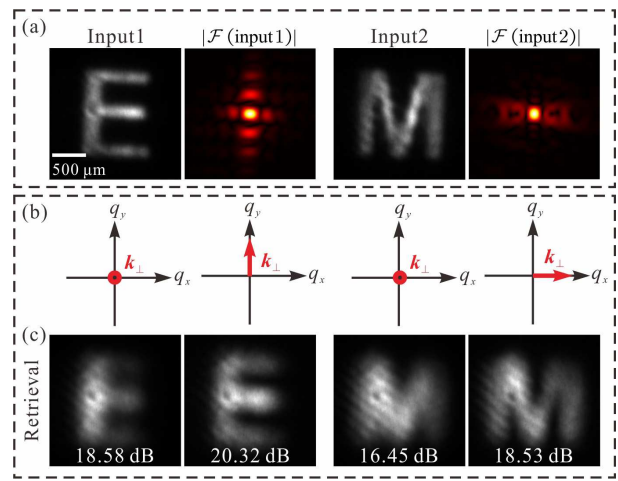}
\caption{Storage of two letters ``E'' and ``M''. (a) Input patterns and their corresponding spatial frequency (i.e., $|$$\mathcal{F}$(input1)$|$ and $|$$\mathcal{F}$(input2)$|$). (b) Depictions of four cases of customized $\textbf{k}_{\perp}$ in the Fourier space. Specific parameters used from left to right: $\alpha=0$ and $\beta \approx 0$ mrad, $\alpha=\pi/2$ and $\beta \approx 0.88$ mrad, $\alpha=0$ and $\beta \approx 0$ mrad, $\alpha=0$ and $\beta \approx 0.86$ mrad. (c) Retrieved patterns corresponding to the four cases shown in (b) with the storage time of $t=5$ $\mu$s. The white numbers indicate the peak signal-to-noise ratio of the retrieved images.}\label{fig:6}
\end{figure}
\section{Wavevector misalignment in improving storage performance of more general images}
As aforementioned, the high SF components determine an optical image's detail. Therefore, the preservation of the high SF components of the retrieved image is of great help to improve the storage fidelity. For a certain image, if its SF mainly distributes along a certain direction, one can design a reasonable $\textbf{k}_{\perp}$ to increase the storage performance.

For instance, we demonstrate the storage of two letters, ``E'' and ``M'', under the collinear and noncollinear cases. The experimentally generated patterns of these two letters and their SF are shown in Fig. \ref{fig:6}(a), from which we can see that the SF of ``E'' (``M'') mainly distributes along $y$-axis ($x$-axis). We store them both for 5 $\mu$s with $\textbf{k}_{\perp}=0$ and $\textbf{k}_{\perp}$ along $y$-axis ($x$-axis) for ``E'' (``M''), and the storage results are shown in Fig. \ref{fig:6}(b). The peak signal-to-noise ratio (PSNR) of each image with respect to the input image is also given to quantitatively compare the results. As can be seen that the retrieved images' qualities under the noncollinear cases are much better than those under the collinear cases and the PSNR values increase by about 2 dB. The method proposed here has greater superiority and convenience over the previous schemes.

\section{conclusions}
In conclusion, by employing the light-storage technique, we experimentally investigate the angular dependency of spatial frequency modulation on the optical images in thermal atoms. It turns out that when the wavevector difference $\textbf{k}$ between the probe and control beams is impinged onto thermal atoms, the effect of atomic diffusion on the probe field is transformed from a low-pass filter into a band-pass filter in the SF domain. Intriguingly, such filter's center frequency is located at $\textbf{k}$,  which is an easily adjustable parameter in experiment. We further demonstrate that for images whose SF mainly distributes along a certain direction, its storage performance can be greatly improved by customizing $\textbf{k}$ in this direction. Our demonstrations will be useful for multimode quantum memory, all-optical structured field manipulation, and imaging through diffusive media, etc.

\begin{acknowledgments}
This work is supported by National Natural Science Foundation of China (NSFC) (12104358, 11774286, and 92050103). 
\end{acknowledgments}

\section*{Appendix A: experimental arrangement} \label{Appendix A}
\begin{figure}[tp]
\centering\includegraphics[width=\linewidth]{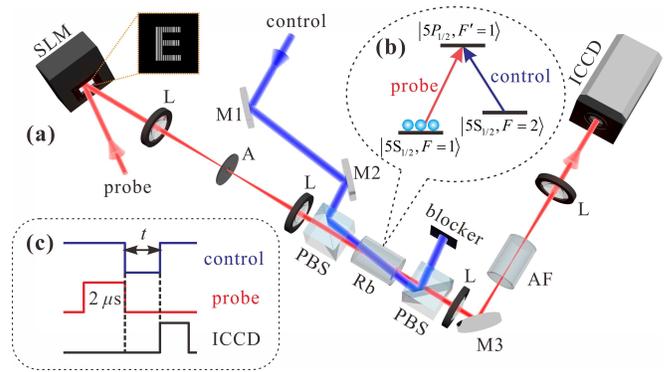}
\caption{(a) Experimental setup. SLM: spatial light modulator, L: lens with a focal length of $f=500$ mm, A: aperture, PBS: polarizing beam splitter, AF: atomic filter, ICCD: intensified charge-coupled device, M1-M3: mirrors. (b) Energy level structure. (c) Time sequence.}\label{fig:7}
\end{figure}

The schematic diagram of the experimental arrangement is illustrated in Fig. \ref{fig:7}(a). A horizontally polarized probe is shaped to a desired image or an LG mode via a spatial light modulator (SLM) with the method in Ref. \cite{Bolduc:13}. We note that the SLM here only operates as an amplitude modulation device for the generation of diverse optical images, which is equivalent to binary image masks but possesses superior adjustment flexibility. The shaped probe is then imaged to the center of the atomic vapor cell by a 4$f$ imaging system. Meanwhile, a vertically polarized control beam intersects with the probe at the center of the atomic vapor cell. The wavevector of the control beam is adjusted by two mirrors (M1 and M2), while the wavevector of the probe remains unchanged. After the atomic cell, a polarizing beam splitter (PBS) together with an atomic filter (AF), which is an 80 $^{\circ}$C atomic cell with all the atoms populated on $|5S_{1/2},F=2\rangle$, are used to filter out the control beam. The retrieved probe is further imaged onto an intensified charge-coupled device (ICCD, Andor iStar 334T) camera by another 4$f$ system.

As depicted in Fig. \ref{fig:7}(b), the control (795 nm, Toptica DL) and probe (795 nm, Toptica DL pro) beams are resonantly coupling to $|5S_{1/2},F=2\rangle\rightarrow|5P_{1/2},F^{\prime}=1\rangle$ and $|5S_{1/2},F=1\rangle\rightarrow|5P_{1/2},F^{\prime}=1\rangle$ transitions respectively, and they are phased-locked through an offset phase lock servo (Vescent D2-135). Moreover, the two beams are chopped into optical pulses (shown in Fig. \ref{fig:7}(c)) by two acousto-optic modulators (AOMs). The shutter of the ICCD synchronizes with the retrieval process.

\section*{Appendix B: detailed simulations of the atomic filtering effect}\label{Appendix B}
\begin{figure*}[tp]
\centering\includegraphics[width=10cm]{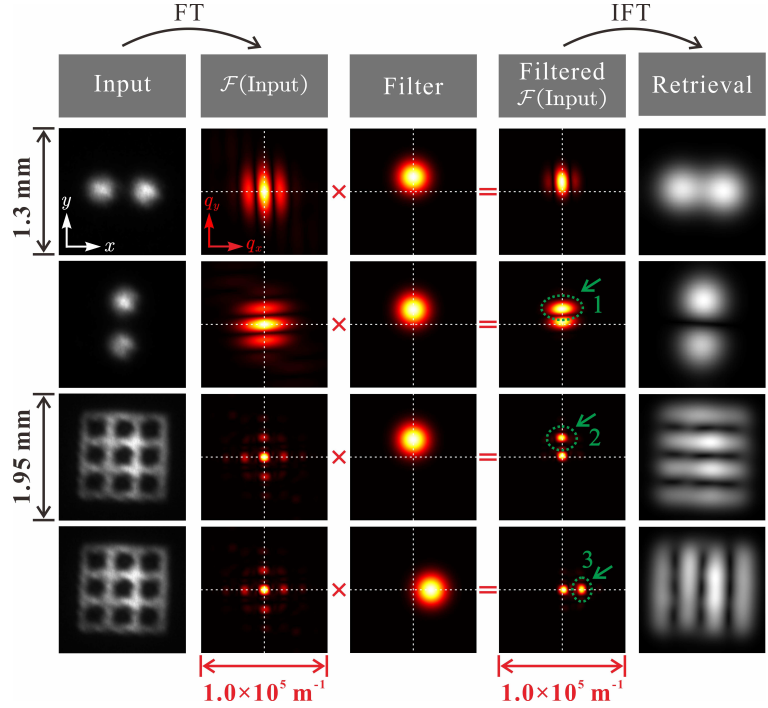}
\caption{Simulations of the atomic band-pass filter acting on the double-petal images and the grid patterns. For the two double-petal images, both of the $\textbf{k}_{\perp}$ are set to be along $y$-axis and the angular deviations are $\beta=1.48$ mrad (same as Fig. \ref{fig:2}). For the grid pattern, $\textbf{k}_{\perp}$ is customized along $y$-axis with $\beta=1.76$ mrad (the third row) and along $x$-axis with $\beta=1.76$ mrad (the fourth row). Columns 1 to 5 are the images to be stored, their spatial spectra, band-pass filtering responses of the atomic vapor, the spatial spectra after filtering and the corresponding retrieved images, respectively. All the spatial spectra shown here have the same size of $10^{5}$ m$^{-1}$ $\times$ $10^{5}$ m$^{-1}$.}\label{fig:8}
\end{figure*}

Figure \ref{fig:8} shows the detailed simulation process of the atomic filter acting on two double-petal images and two grid patterns. The first column shows the input images, the second to fourth columns reveal the filtering operations in the frequency space, and the fifth column shows the retrieved images. In these four cases, the storage time is set to 3 $\mu$s, rendering a filter bandwidth of $1/\sqrt{2Dt}\approx 0.82\times 10^{4}$ m$^{-1}$ (the cut-off frequency of a Gaussian filter is usually defined by the standard deviation in the frequency domain). For the two double-petal images, $\textbf{k}_{\perp}$ is always set to be along $y$-axis and the angular deviation remains $\beta=1.48$ mrad (same as the non-collinear case in Fig. \ref{fig:2}). Under this condition, the center of the filter shifts along $q_{y}$-axis by a distance of $|\textbf{k}_{\perp}|\approx 1.17\times10^{4}$ m$^{-1}$. This shift only enhances the transmissivity of a certain high-frequency component on $q_{y}$-axis, e.g., the spatial frequency component indicated by the green circle (labeled as `1') in Fig. \ref{fig:8}, which locates near $q_{y}=1.27 \times 10^{4}$ m$^{-1}$ ($q_{x}=0$) and determines the details of the vertical double-petal image. The amount of transmissivity enhancement mainly depends on the orientation and the magnitude of $\textbf{k}_{\perp}$, as illustrated in Fig. \ref{fig:2} and Fig. \ref{fig:3}. It is worth reminding here that, first, the low frequency component of the image is generally too high to be completely filtered out. And second, for an image (or a real signal), its Fourier transform is symmetric about the origin, and the coefficients of the positive and negative frequencies are complex conjugates, thus the filtering effect of $\textbf{k}_{\perp}$ shifting along a certain direction or its inversely symmetric direction is equivalent.

For the grid patterns with two designed $\textbf{k}_{\perp}$ along $y$-axis and $x$-axis in Fig. \ref{fig:4}, the corresponding filtering details are shown in the third and fourth rows of Fig. \ref{fig:8}. In these two cases, the centers of the filters shift by $|\textbf{k}_{\perp}|\approx 1.39\times10^{4}$ m$^{-1}$ along $q_{y}$ and $q_{x}$ axes respectively, which locate around the high frequency regions (marked as green circles `2' and `3', corresponding to $q_{y}= 1.5\times10^{4}$ m$^{-1}$ ($q_{x}=0$) and $q_{x}= 1.5\times 10^{4}$ m$^{-1}$ ($q_{y}$=0)). Compare with the low-pass filter under the collinear configuration, this shift enhance the transmissivity of the high frequency component around the region `2' or `3' by about 5 times, while the other high frequency components are significantly weakened. As a result, the dominant high frequency component combined with the retained low frequency component enable the reconstruction of the image ``$\rotatebox{90}{$||||$}$'' or ``$||||$'', as shown in Fig. \ref{fig:8}. These theoretical simulations are in good agreement with our experimental results.

\section*{Appendix C: analytical solution for storing two-point image}\label{Appendix C}
Equation (\ref{Eq. 5}) concisely illustrate the influence of the orientation of $\textbf{k}_{\perp}$ and the diffusion direction on the interference outcomes. For a more quantitative analysis, we can obtain a specific solution for the stored two-petal image by solving the diffusion equation. To facilitate the solution, the two petals are simplified into two points (located at $\textbf{r}=\textbf{r}_{1}$ and $\textbf{r}=\textbf{r}_{2}$ ), which can be expressed by the Dirac $\delta$ function with the form
\begin{equation}
\psi_{points}(\textbf{r})=\delta({\textbf{r}-\textbf{r}_{1}})+\delta({\textbf{r}-\textbf{r}_{2}}).
\tag{A1}
\label{eq.A1}
\end{equation}

Substituting Eq. \ref{eq.A1} into Eq. \ref{Eq. 3}, and using the convolution theorem we obtain the field distribution of the two-point image at storage time $\tau$,
\begin{equation}
\begin{split}
\psi_{points}(\textbf{r},\tau)=\int \delta({\tilde{\textbf{r}}-\textbf{r}_{1}})e^{-i\textbf{k}_{\perp}\cdot(\textbf{r}-\tilde{\textbf{r}})-(\textbf{r}-\tilde{\textbf{r}})/(4D\tau)}d\tilde{\textbf{r}}\\ +\int \delta({\tilde{\textbf{r}}-\textbf{r}_{2}})e^{-i\textbf{k}_{\perp}\cdot(\textbf{r}-\tilde{\textbf{r}})-(\textbf{r}-\tilde{\textbf{r}})/(4D\tau)}d\tilde{\textbf{r}}. 
\end{split}
\tag{A2}
\label{eq.A2}
\end{equation}

With the sampling property of the $\delta$ function, Eq. \ref{eq.A2} is simplified as
\begin{equation}
\begin{split}
\psi_{points}(\textbf{r},\tau)=e^{-i\textbf{k}_{\perp}\cdot(\textbf{r}-\textbf{r}_{1})-(\textbf{r}-\textbf{r}_{1})^{2}/(4D\tau)}\\+e^{-i\textbf{k}_{\perp}\cdot(\textbf{r}-\textbf{r}_{2})-(\textbf{r}-\textbf{r}_{2})^{2}/(4D\tau)},
\end{split}
\tag{A3}
\label{eq.A3}
\end{equation}
where the first and second terms correspond to the first and second points respectively. It can be found that the retrieved two points carry the phase related to $\textbf{k}_{\perp}$, and their phase difference is $\phi_{points}=\textbf{k}_{\perp}\cdot(\textbf{r}_{1}-\textbf{r}_{2})$. When $\textbf{k}_{\perp}=0$ or $\textbf{k}_{\perp}\perp (\textbf{r}_{1}-\textbf{r}_{2})$, $\phi_{points}=0$, meaning that no destructive interference occurs (see Fig. 2(b1)-(b4) and Fig. 2(c1)); Otherwise, $\phi_{points}\neq 0$, and in this case it is possible to satisfy the perfect destructive interference condition, i.e., $\phi_{points}=\pi$. Ulteriorly, since $\phi_{points}=|\textbf{k}_{\perp}|\cdot|\textbf{r}_{1}-\textbf{r}_{2}|\,{\rm sin}\,\theta$ with $\theta$ the rotation angle of the two points around their center (see Fig. \ref{fig:2}), we can deduce that when $\theta=\pi/2$ (or $\textbf{k}_{\perp}\parallel (\textbf{r}_{1}-\textbf{r}_{2})$), the required angle deviation is the smallest to satisfy the destructive interference condition (see Fig. \ref{fig:2}(c4)). Furthermore, when $\theta$ and $|\textbf{r}_{1}-\textbf{r}_{2}|$ are fixed, $\phi_{points} \propto |\textbf{k}_{\perp}|\propto \beta$. Hence there is only one $\beta$ within a phase period to satisfy $\phi_{points}=\pi$ and a larger or a smaller $\beta$ could both reduce the visibility, as revealed by Fig. \ref{fig:3}.

\bibliography{reference}

\end{document}